# Shannon Entropic Entanglement Criterion in the Simple Harmonic Oscillator

**Gardo Blado[1], Amanda Johnson[2,1], Otto Gadea[4,3,1], Francisco Herrera[1],**

[1]*Physics Discipline,* [2]*Department of Biology,* [3]*Chemistry Discipline*

*College of Science and Engineering*

*Houston Baptist University*

*7502 Fondren Rd., Houston, Texas, U.S.A*

[4] *Earth and Atmospheric Science Department*

*3507 Cullen Blvd*

*University of Houston*

*Houston, Texas, U.S.A*

## Abstract

We apply Walborn's[1] Shannon Entropic Entanglement Criterion (SEEC), to simple harmonic oscillators, a system typically discussed in undergraduate quantum mechanics courses. In particular, we investigate the entanglement of a system of coupled harmonic oscillators. A simple form of entanglement criterion in terms of their interaction is found. It is shown that a pair of interacting ground state coupled oscillators are more likely to be entangled for weaker coupling strengths while coupled oscillators in excited states entangle at higher coupling strengths. Interacting oscillators are more likely to be entangled as their interaction increases.

Key words: Shannon entropy, entanglement, harmonic oscillator

## 1. Introduction

In quantum mechanics entanglement can be thought as a correlation between different states with more than one degree of freedom or between particles [2]. It is a field of active research both theoretically and experimentally. The concept of entanglement is usually introduced via the spin entanglement of fundamental particles [2]. However, entanglement in terms of continuous variables $x$ (or $q$) and $p$. [1, 3-8] have also been studied.

Various methods of quantifying entanglement [9] have been used such as the positive partial transpose criterion [3], inseparability criterion using variance form of local uncertainty principle [6], and using the entropic functions such as the Shannon, Renyi and von Neumann entropies [1, 4, 10, 11]. The present work will use the entropic uncertainty relation (EUR) statement of entanglement. We specifically choose the Shannon EUR because of its fundamental nature and due to the vast application of the Shannon entropy [12, 13].

As mentioned in [2] it is important to introduce the concept of entanglement at the undergraduate level because of its applications to the emerging field of quantum information. In addition, its discussion leads to a better understanding of quantum mechanical ideas which helps to prevent common misconceptions in quantum physics. The authors also note that the concept of entanglement fascinates physics and non-physics students alike. Schroeder [2] showed that entanglement can be introduced using the commonly discussed wavefunctions (which involve continuous variables) in any introductory class in quantum physics, beyond the typical use of spin entanglement. This paper is meant to further expose undergraduates to entanglement using topics studied in undergraduate physics such as the simple harmonic oscillator, Hermite polynomials and the hypergeometric functions.

The quantum mechanical simple harmonic oscillator is a common topic discussed in undergraduate quantum mechanics classes. In this paper, we apply the Shannon entropic uncertainty criterion to a system of coupled harmonic oscillators. The entanglement in a system of coupled harmonic oscillators has been studied thoroughly in the literature. For instance, the von Neumann and Renyi entanglement entropies and Schmidt modes have been used to quantify the degree of entanglement for different energies/temperatures [14-16]. Entanglement dynamics (time-dependent interaction strength, sudden death, revival, etc.) have been examined under varying conditions such as initial temperature, damping factors and squeezing of the system and/or its surrounding environment using coupled harmonic oscillators [17-20]. Other applications of the coupled harmonic oscillator among many others include the following. A mathematical formalism that unifies quantum mechanics and special relativity for Lorentz-covariant states was developed through the group symmetries of coupled harmonic oscillators to show that the quark model and Feynman's parton picture can work together to explain the properties of hadrons in high-energy laboratories [21-25]. In [26], the authors used a system of quantum mechanical coupled oscillators to study the effects on a system of interest which can be measured (first oscillator) when one sums over variables of the external system whose variables are not measured (second oscillator which then is "Feynman's rest of the universe") by calculating the entropy using the density matrix formalism.

In a recent paper, the Shannon entropy has been calculated for a single D-dimensional simple harmonic oscillator [13]. The present work calculates the Shannon entropy for a bipartite system of coupled harmonic oscillators to quantify the entanglement of the ground and excited states of this system.

The paper is organized as follows. In section 2, we review the coordinate space and momentum space solutions of a system of coupled oscillators. The new results using the Shannon entanglement criterion are derived in section 3. We give our conclusions in section 4.

## 2. Coupled SHO

In this paper, we consider two masses whose interaction is given by the simple harmonic oscillator (SHO) Hamiltonian

*Equation 1:* $\mathcal{H} = \frac{P_1^2}{2m_1} + \frac{P_2^2}{2m_2} + \frac{A}{2}X_1^2 + \frac{B}{2}X_2^2 + \frac{C}{2}X_1X_2$

Equation 1 which describes a coupled simple harmonic oscillator has a wide variety of applications in physics [18, 26-27].

Let us work out carefully how to transform the Hamiltonian of the SHO, to involve dimensionless variables to enable us to calculate the Shannon entropies. Consider the Hamiltonian $\mathcal{H}$ of one oscillator given by

Equation 2: $\mathcal{H} = \frac{P^2}{2m} + \frac{1}{2}kX^2 = \frac{P^2}{2m} + \frac{1}{2}m\omega^2 X^2$

where $\omega = \sqrt{\frac{k}{m}}$ with energies $E_n = \left(n + \frac{1}{2}\right)\hbar\omega$, $n = 0, 1, 2, \ldots$ in which $[X, P] = i\hbar$. Dividing the Hamiltonian by $\hbar\omega$, we get

Equation 3: $H = \frac{1}{2}(p)^2 + \frac{1}{2}(y)^2$

where we have the dimensionless variables $H \equiv \frac{\mathcal{H}}{\hbar\omega}$, $p \equiv \frac{P}{(m\hbar^2 k)^{1/4}} = \frac{P}{\sqrt{m\hbar\omega}}$, and $y \equiv \frac{X}{(\hbar^2/(mk))^{1/4}} = \frac{X}{\sqrt{\hbar/(m\omega)}}$. Clearly, $p = -i\frac{d}{dy}$ and one can easily show that

Equation 4: $[y, p] = i$.

In the literature, the above results are often equivalently obtained by setting $\hbar$, $m$, $k$ and $\omega$ to 1.

Let us reconsider the Hamiltonian of two masses interacting as coupled harmonic oscillators in Equation 1, $\mathcal{H} = \frac{P_1^2}{2m_1} + \frac{P_2^2}{2m_2} + \frac{A}{2}X_1^2 + \frac{B}{2}X_2^2 + \frac{C}{2}X_1 X_2$ with $P_j = -i\hbar\frac{d}{dX_j}$. One can diagonalize this to give [14, 26]

Equation 5: $\mathcal{H} = \frac{1}{2M}\left(\left[P_1'\right]^2 + \left[P_2'\right]^2\right) + \frac{1}{2}K\left(e^{2\eta}\left[X_1'\right]^2 + e^{-2\eta}\left[X_2'\right]^2\right)$

where $M = \sqrt{m_1 m_2}$, $P_j' = -i\hbar\frac{d}{dX_j'}$, $j = 1, 2$, $K = \sqrt{AB - C^2/4}$,

Equation 6: $e^{2\eta} = \frac{A + B + \frac{A-B}{|A-B|}\sqrt{(A-B)^2 + C^2}}{\sqrt{4AB - C^2}}$

and

Equation 7: $X_1' = X_1 \cos\alpha - X_2 \sin\alpha$, $X_2' = X_1 \sin\alpha + X_2 \cos\alpha$

with

Equation 8: $\tan 2\alpha = \frac{C}{B - A}$.

We carefully changed our notation with the primed coordinates to explicitly exhibit the transformations to diagonalize the original Hamiltonian $\mathcal{H}$. The details can be found in [26].

Similar to Equation 3 which came from Equation 2, we can rewrite Equation 5 in terms of dimensionless variables $p_1, p_2, y_1$, and $y_2$.

Equation 9: $H = \frac{1}{2}p_1^2 + \frac{1}{2}p_2^2 + \frac{1}{2}e^{2\eta}y_1^2 + \frac{1}{2}e^{-2\eta}y_2^2$

with $[y_j, p_k] = i\delta_{jk}$, $H = \frac{\mathcal{H}}{\hbar\omega}$, $p_j \equiv \frac{P_j'}{(M\hbar^2 K)^{1/4}}$, $y_j \equiv \frac{X_j'}{(\hbar^2/(MK))^{1/4}}$ and $\omega = \sqrt{\frac{K}{M}}$, or equivalently as mentioned before, let $\hbar$, $M$, $K$ and $\omega$ equal 1. Defining the dimensionless variable $x_j \equiv \frac{X_j}{(\hbar^2/(MK))^{1/4}}$, Equation 7 becomes

Equation 10: $y_1 = x_1 \cos\alpha - x_2 \sin\alpha$, $y_2 = x_1 \sin\alpha + x_2 \cos\alpha$.

We can infer from reference [14] the solution of the Schrodinger equation with the Hamiltonian given by Equation 9.

Equation 11: $\Psi_{nm} = \psi_n \psi_m = c_n^{(1)} c_m^{(2)} \exp\left(-\frac{e^\eta}{2} y_1^2\right) \exp\left(-\frac{e^{-\eta}}{2} y_2^2\right) H_n(e^{\eta/2} y_1) H_m(e^{-\eta/2} y_2)$

where

Equation 12: $c_n^{(1)} = \frac{1}{\sqrt{\sqrt{\pi} n! 2^n}}$ and $c_m^{(2)} = \frac{1}{\sqrt{\sqrt{\pi} m! 2^m}}$.

The $H_n$ and $H_m$ are the Hermite polynomials. The corresponding eigenvalues are

Equation 13: $E_{nm} = e^\eta \left(n + \frac{1}{2}\right) + e^{-\eta} \left(m + \frac{1}{2}\right)$.

Similarly, one can either solve the momentum space wave functions by finding the Fourier transform of Equation 11 or simply replacing $e^{\eta/2} y_1$ by $e^{-\eta/2} p_1$ and $e^{-\eta/2} y_2$ by $e^{\eta/2} p_2$ in Equation 11 similar to [13].

Equation 14: $\Phi_{nm} = \varphi_n \varphi_m = c_n^{(1)} c_m^{(2)} \exp\left(-\frac{e^{-\eta}}{2} p_1^2\right) \exp\left(-\frac{e^\eta}{2} p_2^2\right) H_n(e^{-\eta/2} p_1) H_m(e^{\eta/2} p_2)$.

## 3. Shannon Entropic Entanglement Criterion for SHO

EUR were introduced as an alternative way to express the Heisenberg uncertainty principle, to address some shortcomings of the variance statement $\Delta x \Delta p \geq \hbar/2$ [12, 28]. One of the advantages of using the EUR is that it is an excellent mathematical framework to quantify uncertainties for correlated systems such as entangled systems [28]. The first measure of uncertainty in terms of entropy given by Hartley (which he called "information") was $U(n) = k \log n$ where $U$ is the uncertainty of $n$ outcomes of a random experiment. For the case in which the probabilities of the outcomes are unequal, this is generalized to $U_i = k \sum_i^n P_i \log(1/P_i)$ where $P_i$ is the probability of the *i*th outcome. For $k = 1$, we get the Shannon entropy $S = -\sum_{i=1}^n P(x_i) \ln P(x_i)$ with $x$ as a discrete random variable. It can be shown that the preceding equation can be generalized to

*Equation 15:* $S = -\int_{-\infty}^{\infty} p(x) \ln p(x) \, dx$

for a continuous variable $x$ with $p(x)$ as the probability density function. We will refer to Equation 15 as the Shannon entropy. We refer the reader to [12] for more details.

To study the entanglement of coupled harmonic oscillators using the Shannon entropy, consider a bipartite system which we label as system 1 and system 2. In the succeeding discussions, subscripts 1 and 2 correspond to systems 1 and 2 respectively. As in [1, 10], we define the dimensionless variables as,

*Equation 16:* $x_\pm = x_1 \pm x_2; \; p_\pm = p_1 \pm p_2$.

with $[x_j, p_k] = i\delta_{jk}$ as in Equation 4. A pure state of the system is described by the wavefunctions $\Psi(x_1, x_2) = \psi_1(x_1)\psi_2(x_2)$ in coordinate space and $\mathcal{P}(p_1, p_2) = \varphi_1(p_1)\varphi_2(p_2)$ in momentum space. A change in variables using Equation 16, gives $\Psi(x_+, x_-) = \frac{1}{\sqrt{2}} \psi_1\left(\frac{x_+ + x_-}{2}\right) \psi_2\left(\frac{x_+ - x_-}{2}\right)$; $\mathcal{P}(p_+, p_-) = \frac{1}{\sqrt{2}} \varphi_1\left(\frac{p_+ + p_-}{2}\right) \varphi_2\left(\frac{p_+ - p_-}{2}\right)$.

The Shannon entropic entanglement criterion is given by

*Equation 17:* $H[w_\pm] + H[v_\mp] < \ln(2\pi e)$ or $H[w_\pm] + H[v_\mp] - \ln(2\pi e) < 0$

where the Shannon entropies are given by

*Equation 18:* $H[w_\pm] = -\int_{-\infty}^{\infty} dx_\pm w_\pm(x_\pm) \ln\left(w_\pm(x_\pm)\right); \; H[v_\pm] = -\int_{-\infty}^{\infty} dp_\pm v_\pm(p_\pm) \ln\left(v_\pm(p_\pm)\right)$

with

*Equation 19:* $w_\pm(x_\pm) = \int_{-\infty}^{\infty} dx_\mp |\Psi(x_+, x_-)|^2 = \frac{1}{2} \int_{-\infty}^{\infty} dx_\mp |\psi_1|^2 |\psi_2|^2$; $v_\pm(p_\pm) = \int_{-\infty}^{\infty} dp_\mp |\mathcal{P}(p_+, p_-)|^2 = \frac{1}{2} \int_{-\infty}^{\infty} dp_\mp |\varphi_1|^2 |\varphi_2|^2$.

Strictly speaking in Equation 6, the case A = B is not defined. However, to facilitate a simpler calculation, we will look at the case in which $A \approx B$. Let us look at the two cases in which $A \to B$ for values A > B and for values A < B. For the case A > B, Equation 6 becomes $e^{2\eta} = \frac{A+B+\sqrt{(A-B)^2+C^2}}{\sqrt{4AB-C^2}}$, and $A \to B$ yields

Equation 20: $e^{2\eta} \approx \sqrt{\frac{2A+|C|}{2A-|C|}} > 1$

for non-zero C. This implies that $\eta > 0$. For the case A < B, Equation 6 becomes $e^{2\eta} = \frac{A+B-\sqrt{(A-B)^2+C^2}}{\sqrt{4AB-C^2}}$,

and $A \to B$ yields $e^{2\eta} \approx \sqrt{\frac{2A-|C|}{2A+|C|}} < 1$ for non-zero C. This implies that $\eta < 0$. As will be seen later in this section, entanglement occurs only for $\eta > 0$. Let us then consider the case when A > B. Since the value of $\eta$ in section 2 above is independent of the sign of C, let us assume for the moment that $C < 0$[1].

Equation 8 gives $\alpha = 45^0$ and $y_1 = \frac{x_1}{\sqrt{2}} - \frac{x_2}{\sqrt{2}}$, $y_2 = \frac{x_1}{\sqrt{2}} + \frac{x_2}{\sqrt{2}}$ from Equation 7. From Equation 16, we get

Equation 21: $y_1 = x_-/\sqrt{2}$, $y_2 = x_+/\sqrt{2}$

and similarly

Equation 22: $p_1 = p_-/\sqrt{2}$, $p_2 = p_+/\sqrt{2}$

From Equation 21 and Equation 22 we can rewrite the solutions in Equation 11 and Equation 14 as

Equation 23: $\Psi_{nm} = c_n^{(1)} c_m^{(2)} \exp\left(-\frac{e^\eta}{4} x_-^2\right) \exp\left(-\frac{e^{-\eta}}{4} x_+^2\right) H_n\left(\frac{e^{\eta/2}}{\sqrt{2}} x_-\right) H_m\left(\frac{e^{-\eta/2}}{\sqrt{2}} x_+\right)$

and

Equation 24: $\Phi_{nm} = c_n^{(1)} c_m^{(2)} \exp\left(-\frac{e^{-\eta}}{4} p_-^2\right) \exp\left(-\frac{e^\eta}{4} p_+^2\right) H_n\left(\frac{e^{-\eta/2}}{\sqrt{2}} p_-\right) H_m\left(\frac{e^{\eta/2}}{\sqrt{2}} p_+\right)$

Let us now outline the calculation of the Shannon entropic entanglement criterion by calculating the function

Equation 25: $f(\eta) = H[w_-] + H[v_+] - \ln(2\pi e)$.

Using instead $H[w_+] + H[v_-] - \ln(2\pi e)$ yields the same results.

We start with $H[w_-]$. Letting

Equation 26: $\begin{cases} a)\ t \equiv e^{\eta/2}/\sqrt{2} \\ b)\ z_1 \equiv tx_- = \left(e^{\eta/2}/\sqrt{2}\right)x_- \\ c)\ z_2 \equiv x_+/(2t) = \left(e^{-\eta/2}/\sqrt{2}\right)x_+ \end{cases}$

---

[1] The case C > 0 results in $\alpha = -45^0$ with the same qualitative results. In providing the details, we chose to use C<0 to get the $\alpha = 45^0$ instead which has been discussed in many of the applications mentioned in [26] and which makes the calculations more straightforward without bothering about the extra negative sign introduced by $\alpha = -45^0$.

Equation 23 becomes (note Equation 11 too)

*Equation 27:* $\Psi_{nm} = \psi_n \psi_m = c_n^{(1)} c_m^{(2)} e^{-(z_1^2/2)} e^{-(z_2^2/2)} H_n(z_1) H_m(z_2)$

From Equation 19, we get $w_- = \frac{1}{2}\int_{-\infty}^{\infty} dx_+ |\psi_n|^2 |\psi_m|^2$ and with the change in variables in Equation 26, we get using Equation 27

*Equation 28:* $w_- = q_{nm} e^{-z_1^2} H_n^2(z_1)$ with $q_{nm} = \frac{t I_0}{\pi n! m! 2^n 2^m}$

where the integral

*Equation 29:* $I_0 \equiv \int_{-\infty}^{\infty} e^{-z_2^2} H_m^2(z_2)\, dz_2$.

From Equation 18, $H[w_-] = -\int_{-\infty}^{\infty} dx_- w_- \ln(w_-)$. Using Equation 28 and expanding gives

*Equation 30:* $H[w_-] = -\frac{1}{t} q_{nm}\{(\ln q_{nm}) I_1 + I_2 + I_3\}$

where

*Equation 31:*
$\begin{cases} a)\ I_1 \equiv \int_{-\infty}^{\infty} e^{-z_1^2} H_n^2(z_1)\, dz_1 \\ b)\ I_2 \equiv -\int_{-\infty}^{\infty} z_1^2 e^{-z_1^2} H_n^2(z_1)\, dz_1 \\ c)\ I_3 \equiv \int_{-\infty}^{\infty} e^{-z_1^2} H_n^2(z_1) \ln(H_n^2(z_1))\, dz_1 \end{cases}$.

From [13], we get analytic expressions for the integrals in Equation 29 and Equation 31.

*Equation 32:*
$\begin{cases} a)\ I_0 = 2^m m! \sqrt{\pi} \\ b)\ I_1 = 2^n n! \sqrt{\pi} \\ c)\ I_2 = -2^n n! \sqrt{\pi}\, (n + 1/2) \\ d)\ I_3 = 2^n n! \sqrt{\pi} \ln(2^{2n}) - 2 \sum_{k=1}^{n} V_n(x_{n,k}) \end{cases}$

where $x_{n,k}$ are the roots of $H_n(z_1)$ and $V_n$, called the logarithmic potential of the Hermite polynomial $H_n$, is given by $V_n(x_{n,k}) = 2^n n! \sqrt{\pi} \left[\ln 2 + \frac{\gamma}{2} - x_{n,k}^2\ {}_2F_2\left(1,1;\frac{3}{2}, 2; -x_{n,k}^2\right) + \frac{1}{2} \sum_{i=1}^{n} \binom{n}{k} \frac{(-1)^k 2^k}{k}\ {}_1F_1\left(1;\frac{1}{2}; -x_{n,k}^2\right)\right]$ with $\gamma \approx 0.577$ is the Euler constant and ${}_1F_1$ and ${}_2F_2$ are the hypergeometric functions.

Next we calculate $H[v_+]$. Letting (similar to Equation 26)

*Equation 33:*
$\begin{cases} a)\ t \equiv e^{\eta/2}/\sqrt{2} \\ b)\ p_1 \equiv p_-/(2t) = (e^{-\eta/2}/\sqrt{2}) p_- \\ c)\ p_2 \equiv t p_+ = (e^{\eta/2}/\sqrt{2}) p_+ \end{cases}$

Equation 24 becomes (note Equation 14 too)

*Equation 34:* $\Phi_{nm} = \varphi_n \varphi_m = c_n^{(1)} c_m^{(2)} e^{-(\wp_1^2/2)} e^{-(\wp_2^2/2)} H_n(\wp_1) H_m(\wp_2)$

From Equation 19, we get $v_+ = \frac{1}{2} \int_{-\infty}^{\infty} d\wp_- |\varphi_n|^2 |\varphi_m|^2$ and with the change in variables in Equation 33, we get using Equation 34,

*Equation 35:* $v_+ = r_{nm} e^{-\wp_2^2} H_m^2(\wp_2)$ with $r_{nm} = \frac{tJ_0}{\pi n! m! 2^n 2^m}$

where the integral $J_0 \equiv \int_{-\infty}^{\infty} e^{-\wp_1^2} H_n^2(\wp_1) d\wp_1$. From Equation 18, $H[v_+] = -\int_{-\infty}^{\infty} d\wp_+ v_+ \ln(v_+)$.

Using Equation 35 and expanding, we get

*Equation 36:* $H[v_+] = -\frac{1}{t} r_{nm} \{(\ln r_{nm}) J_1 + J_2 + J_3\}$

where $J_1 \equiv \int_{-\infty}^{\infty} e^{-\wp_2^2} H_m^2(\wp_2) d\wp_2$, $J_2 \equiv -\int_{-\infty}^{\infty} \wp_2^2 e^{-\wp_2^2} H_n^2(\wp_2) d\wp_2$ and $J_3 \equiv \int_{-\infty}^{\infty} e^{-\wp_2^2} H_n^2(\wp_2) \ln(H_n^2(\wp_2)) d\wp_2$. With a proper change in variables, one can easily evaluate $J_0, J_1, J_2$, and $J_3$ similar to the evaluation of the integrals in Equation 29, Equation 31 as given by Equation 32. The above discussion of the integrals facilitates the Maple calculation of $f(\eta)$ in Equation 25 using Equation 30 and Equation 36 for different values of $n$ and $m$. In general, the $f(\eta)$ has the form

*Equation 37:* $f(\eta) = \eta_0 - \eta$

where $\eta_0$ is a constant (threshold value) which is the $\eta$-intercept (horizontal axis intercept).

In *Figure 1*, we plot $f(\eta)$ in Equation 25. For instance, $f(\eta) = -\eta$ for $(n = m = 0)$; $f(\eta) \approx 0.541 - \eta$ for $(n = m = 1)$; $f(\eta) \approx 0.852 - \eta$ for $(n = m = 2)$; $f(\eta) \approx 1.07 - \eta$ for $(n = m = 3)$, etc. Similar graphs can be plotted for states with unequal values of $n$ and $m$. From Equation 17, we note that entanglement occurs when $f(\eta) < 0$. The graph shows the strong dependence of entanglement on the quantum numbers $(n, m)$.

For the case we are considering in Equation 20, for no interaction, $C = 0$, implies $\eta = 0$. As mentioned earlier, indeed entanglement occurs for $\eta > \eta_0 \geq 0$. As expected when there is no interaction ($\eta = 0$), there is no entanglement. A higher value of $|C|$ (and thus $\eta$) implies a stronger coupling. For a particular state (given $n$ and $m$), stronger interactions (higher $\eta$ values) result in higher degrees of entanglement (more negative f(η)). Ground state $(n = m = 0)$ interacting oscillators are entangled for $\eta > \eta_0 = 0$.

For a given $\eta$ (given couplings *A* and *C*) however, oscillators in excited states $(n > 0$ or $m > 0)$ are entangled at a nonzero threshold value of $\eta_0$ in Equation 37. For example, as given above, for the state ($n = m = 3$), the threshold value is $\eta_0 = 1.07$. The next higher energy state from the ground state, n = 1, m = 0 or n = 0, m = 1 give a threshold of $\eta_0 = 0.270$. Excited states generally have a higher threshold than the ground state because as shown in [29] excited states have higher entropy than the ground state. In addition, excited states will need stronger interactions (corresponding to higher $\eta_0$) to lead to entanglement. By using different values of *n* and *m* one can check that the value of the threshold $\eta_0$ is the same for $\psi_{nm}$ and $\psi_{mn}$ although these are different states with different eigenvalues as can be seen from Equation 11 and Equation 13. In other words, the SEEC is symmetric in the quantum numbers *n* and *m*. Figure 2 shows an increase in entropy as shown by an increase in the threshold $\eta_0$ and also exhibits the tendency of entanglement at a higher interaction threshold for increasing quantum numbers (n, m).

## 4. Conclusions

Using the Shannon Entropic Entanglement Criterion (SEEC), we study the entanglement of coupled harmonic oscillators. A simple entanglement criterion is found in terms of the interaction parameter $\eta$, given by $f(\eta) = H[w] + H[v] - \ln(2\pi e) = \eta_0 - \eta < 0$ where $\eta_0$ is a threshold value depending on the state (n,m). It is found that coupled harmonic oscillators in the ground state have a lower threshold for entanglement than excited states. More interaction is needed for the higher energy states to "share" information and hence entangle due to the increased number of possible states. In addition, the SEEC is shown to be symmetric in the quantum numbers n and m. For a given state, increasing the interaction increases the entanglement as expected since a higher interaction results in more information between the coupled oscillators leading to more entanglement. The SEEC can also be applied to quantum wells and preliminary calculations seem to indicate similar results. However, unlike the SHO, in which the integrals involve well known special functions which can be evaluated and expressed in analytic form, all the integrals involved in quantum wells can only be evaluated numerically [12, 30-32].

**FIGURES:**

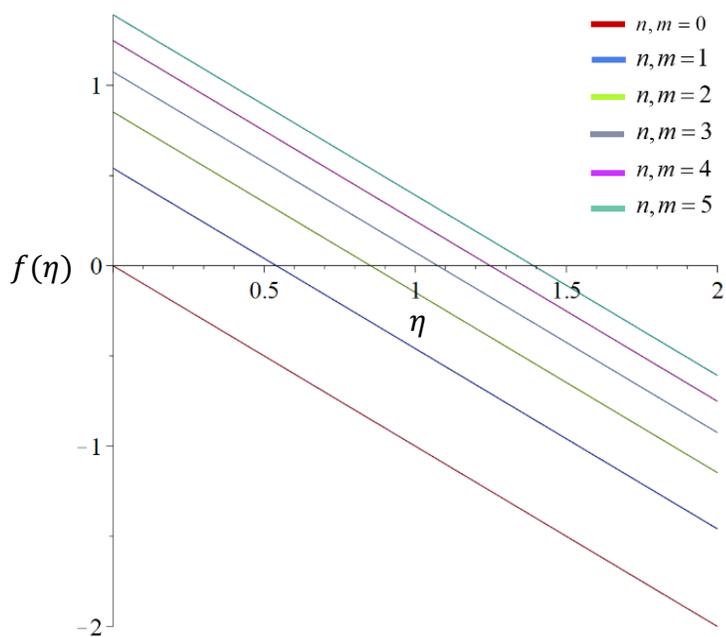

**Figure 1:** Plot of $f(\eta) = H[w_-] + H[v_+] - ln(2\pi e)$ of the ground state and some excited states.

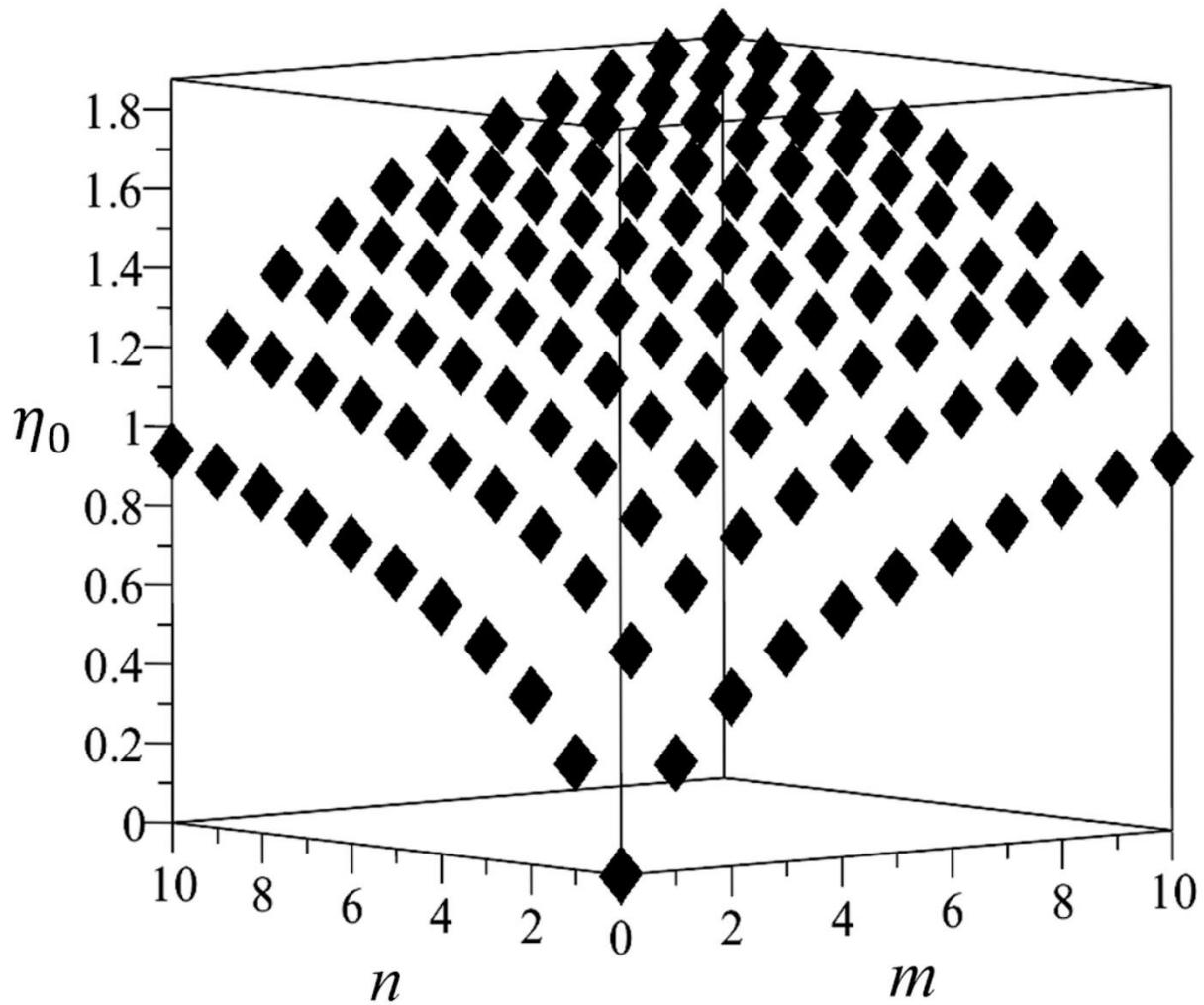

Figure 2: Plot of the threshold values $\eta_0$ given values of n and m

**Acknowledgement:**

OG would like to thank Dr. D. N. Makarov for his kind assistance regarding reference [14].